\documentclass[nolinenumbers,twocolumn]{aastex7}
\usepackage{xcolor}

\shorttitle{Distilled WDs}
\shortauthors{Barrientos et al.}

\begin{document}
\setcounter{footnote}{0}

\title{The Fraction of Distilled White Dwarfs with Long-Lived Habitable Zones}

\author[orcid=0000-0002-6153-9304, sname='Barrientos']{Manuel Barrientos} 
\affiliation{Homer L. Dodge Department of Physics and Astronomy, University of Oklahoma, 440 W. Brooks St., Norman OK, 73019, USA}
\email[show]{mbarrientos@ou.edu}

\author[orcid=0000-0001-6098-2235,sname='Kilic']{Mukremin Kilic} 
\affiliation{Homer L. Dodge Department of Physics and Astronomy, University of Oklahoma, 440 W. Brooks St., Norman OK, 73019, USA}
\email{kilic@ou.edu}

\author[orcid=0000-0002-9632-1436,sname='Blouin']{Simon Blouin}
\affiliation{Department of Physics and Astronomy, University of Victoria, Victoria BC V8W 2Y2, Canada}
\email{sblouin@uvic.ca}

\author[orcid=0000-0001-7294-9766,sname='Hayden']{Michael R. Hayden}
\affiliation{Homer L. Dodge Department of Physics and Astronomy, University of Oklahoma, 440 W. Brooks St., Norman OK, 73019, USA}
\email{mrhayden@ou.edu}

\author[0000-0002-0920-809X,sname='Sharma']{Sanjib Sharma}
\affiliation{Space Telescope Science Institute, 3700 San Martin Drive, Baltimore, MD 21218, USA}
\email{ssharma@stsci.edu}

\author[orcid=0000-0002-0948-4801,sname='Green']{Matthew J. Green}
\affiliation{Homer L. Dodge Department of Physics and Astronomy, University of Oklahoma, 440 W. Brooks St., Norman OK, 73019, USA}
\email{matthew.j.green-2@ou.edu}

\begin{abstract}

After carbon and oxygen, $^{22}$Ne is the most abundant element in white dwarf interiors. As C/O white dwarfs (WDs) crystallize, they are predicted to go through a distillation process in the central layers if they have a sufficiently high $^{22}$Ne mass fraction of $\gtrsim2.5$\%. Observational evidence for distillation comes from an over-density of WDs on the Q-branch in Gaia color-magnitude diagrams, which indicates that $\sim6$\% of massive WDs are delayed in their cooling by as much as $\sim10$ Gyr. However, it is unclear how these stars end up with such a high concentration of $^{22}$Ne and if a significant fraction of the more common average-mass WDs go through distillation. We argue that a significant metal-rich stellar population in the solar neighborhood should lead to distilled WDs, without requiring a binary merger. We use MESA along with the CNO abundances derived from high-resolution spectroscopy of stars included in the Hypatia catalog to predict the $^{22}$Ne mass fraction in their descendant WDs. We find that 0.6-2.5\% of the WDs in the solar neighborhood have sufficient $^{22}$Ne in their interiors to go through multi-Gyr cooling delays, which could significantly inflate their numbers in the observed samples. Hence, $^{22}$Ne distillation and long-lived habitable zones around WDs should be relatively common in the solar neighborhood. We also use a Galactic model to predict the fraction of WDs that go through distillation as a function of Galactocentric distance. The fraction of distilled WDs is $\sim2$-8\% near the Galactic center, and declines steadily toward the outer disk.

\end{abstract}

 
\section{Introduction} 
\label{sec:1}

Stellar evolution leads to C/O core white dwarfs (WDs) for most main-sequence stars unless the progenitor is massive or its evolution is impacted by mass transfer or mergers in binary systems. After carbon and oxygen, the most abundant element in WD interiors is $^{22}$Ne, with a typical mass fraction of $\sim$1.4\% for descendants of solar-composition progenitors \citep{cheng19}. This is because the CNO cycle converts essentially all CNO nuclei into $^{14}$N, which is then transformed into $^{22}$Ne through the chain reaction $^{14}$N($\alpha,\gamma$)$^{18}$F($\beta^{+}$)$^{18}$O($\alpha,\gamma$)$^{22}$Ne during the He-burning stage \citep{deloye02}.

Gravitational energy release from $^{22}$Ne phase separation during core crystallization can have a significant impact on WD cooling \citep{isern91}. \citet{blouin21} calculated the melting curve of C/O/Ne mixtures and found that the solid crystals can be depleted in $^{22}$Ne with respect to the surrounding liquid. Because the crystals are lighter than the surrounding medium, they float up and melt in lower-density regions. This is similar to a distillation process, and it gradually displaces $^{22}$Ne. Gravitational energy release from this mechanism can lead to multi-Gyr cooling delays for WDs with a $^{22}$Ne enrichment of $\gtrsim2.5-3.0\%$ \citep{bedard24,salaris24}.

The evidence that a fraction of WDs suffer from multi-Gyr cooling delays comes from an over-density of these stars on the crystallization sequence \citep{tremblay19} in Gaia color-magnitude diagrams \citep{gaiadr2}. However, the cooling delays from crystallization and C/O phase separation are insufficient to explain the observed over-density \citep[e.g.,][]{montgomery99,kilic2020,kilic2025}. \citet{cheng19} found that about 6\% of massive WDs (M$\gtrsim 1.1$~M$_{\odot}$) in the solar neighborhood suffer from a cooling anomaly that keeps them warm for $\geq$8 Gyr. \citet{blouin21} suggested that $^{22}$Ne distillation in the central layers can provide a plausible explanation for such a large cooling delay, and \citet{bedard24} could explain the luminosity function of massive WDs in Gaia by including the $^{22}$Ne distillation in their evolutionary models. 

Observations of the metal-rich open cluster NGC 6791 provide additional support for the distillation hypothesis. Because $^{22}$Ne originates from the CNO cycle via $^{14}$N during He-burning, its abundance in WDs scales with the progenitor’s initial CNO content and thus with metallicity. The high metallicity of NGC 6791, [Fe/H]$\approx$+0.3, leads to enhanced $^{22}$Ne content in the interiors of its WDs. \citet{salaris24} demonstrate that models including $^{22}$Ne distillation accurately reproduce the faint end of the observed WD luminosity function in this 8.5 Gyr old cluster, whereas models neglecting distillation do not. This relationship between metallicity and $^{22}$Ne content not only motivates the case study of NGC 6791, but also justifies examining the metallicity distribution of stars in the solar neighborhood to estimate what fraction may evolve into $^{22}$Ne-rich WDs.

\begin{figure}
\includegraphics[width=3.4in]{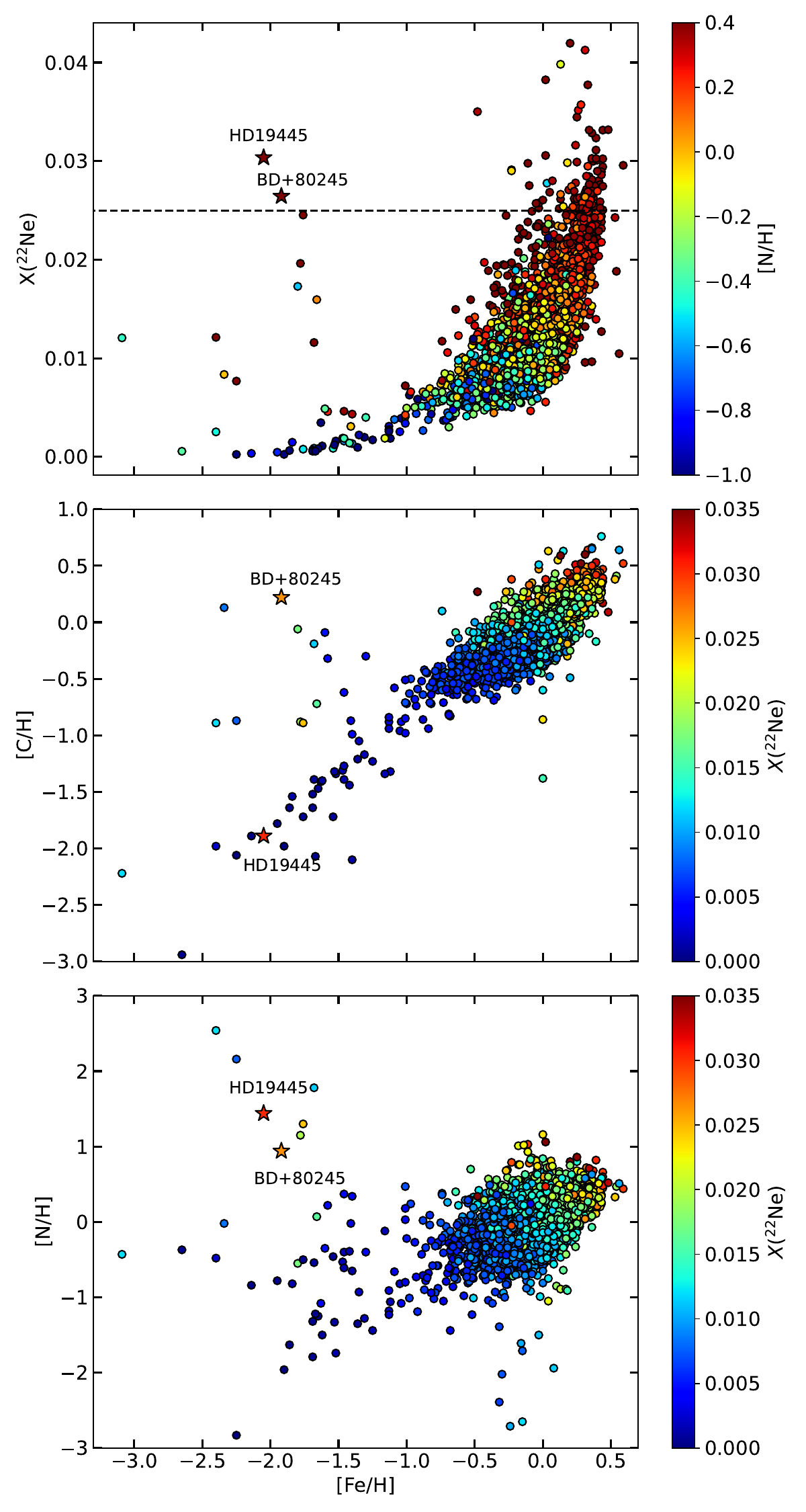}
\caption{{\it Top:} Predicted $^{22}$Ne mass fraction as a function of metallicity for the WD remnants of stars in the Hypatia Catalog \citep{hinkel14}. The colors represent [N/H] for the same stars. {\it Middle and bottom:} C and N abundances of the same stars from the Hypatia Catalog. The color bars represent the predicted neon mass fraction, X($^{22}$Ne), in their remnants. Two chemically peculiar low-metallicity objects, BD+80 245 and HD 19445, are labeled and marked as stars.}
\label{fig:1}
\end{figure}

Distillation has a significant impact on the observational properties of the WD population and their habitable zones. For example, \citet{jewett24} found a remarkably high merger rate of nearly 80\% for 1.1-1.2 $M_{\odot}$ WDs, but they demonstrate that the overabundance of the merger remnant candidates in their sample is likely due to $^{22}$Ne distillation. Since the merger remnants end up with more $^{22}$Ne and $^{26}$Mg \citep{shen23}, they suffer long cooling delays. As a consequence, these WDs remain more luminous for extended periods and are preferentially detected in magnitude-limited surveys, leading to their over-representation in the solar neighborhood. \citet{vanderburg25} modeled the evolution of 0.6, 0.8, and $1.0~M_{\odot}$ WDs with 3\% $^{22}$Ne by mass and found that these stars experience 10, 9, and 6 Gyr cooling delays, respectively. They note that the additional energy source from distillation stabilizes the habitable zone by up to 10 Gyr. \citet{vanderburg25} also highlight the relatively high $^{22}$Ne abundances required for distillation as a problem. They mention this might require either binary mergers or a high primordial abundance of $\alpha$ elements.

In this letter, we investigate the production of $^{22}$Ne in the solar neighborhood WDs by performing MESA simulations based on detailed C, N, O, and [Fe/H] abundance measurements from a sample of $\sim4000$ progenitor stars in the Hypatia Catalog. We estimate the fraction of WDs that go through $^{22}$Ne distillation in the central layers, which leads to significant cooling delays. Other distillation scenarios, such as the formation of a $^{22}$Ne-rich shell, produce smaller cooling delays and therefore are not considered in this study. Section~\ref{sec:2} describes the Hypatia Catalog and outlines the details of our MESA calculations. In Section~\ref{sec:3}, we present the resulting $^{22}$Ne mass fraction for WD progenitors in the solar neighborhood. We expand this analysis by using a Galactic stellar population model to study the distribution of distilled WDs across the Galaxy. Our main conclusions are summarized in Section~\ref{sec:4}.

\section{Hypatia Catalog and MESA Simulations}
\label{sec:2}

\citet{kilic07} found that $\sim5$\% of the stars in the solar neighborhood exhibit metallicities similar to NGC 6791. Consequently, a few percent of their descendant WDs should go through distillation, just like the ones in NGC 6791 \citep{salaris24}. In order to calculate the $^{22}$Ne mass fractions in WDs, we rely on the metallicity measurements for the progenitor stars from the Hypatia Catalog\footnote{\url{https://www.hypatiacatalog.com}}, which includes abundances from over 350 literature sources. Hypatia Catalog is the largest database of high resolution stellar abundances for stars within the solar neighborhood,
and includes stellar elemental abundance measurements for $>10^4$ FGKM-type stars within 500 pc of the Sun \citep{hinkel14,hinkel2016,hinkel2017}. 

We find 4010 stars in Hypatia Catalog with [C/H], [N/H], [O/H], and [Fe/H] measurements. A caveat in using abundance measurements from various sources using different observational setups and abundance analysis techniques is that there are likely systematic uncertainties \citep{hinkel2016}. \citet{hinkel14} found a median spread of 0.11 dex between various measurements for all elements in their catalog of stars. They tried to make the Hypatia Catalog abundance measurements more uniform by putting all of the different measurements on the same solar abundance scale, which results in a median difference of 0.04 dex after re-normalization (see their Section 3.2). However, this is smaller than the typical error bars in most abundance measurements, and therefore, the re-normalization does not have a significant impact on the CNO abundances of the stars in the catalog.

To compute the $^{22}$Ne mass fractions after the He-burning phase in their WD descendants, we calculated evolutionary models of a progenitor main-sequence star of $M=3~M_{\odot}$ by using the test suite template \texttt{make\_co\_wd} of MESA r23.05.1 \citep{paxton19,MESA2023} and adopting the CNO abundances of each star in the Hypatia Catalog. To asses the sensitivity of our results to the initial mass of the progenitor, we repeated the calculations for masses of 1.5, 2, 3, and 4~M$_{\odot}$ assuming solar composition. The determined $^{22}$Ne mass fractions for the WDs varied by no more than 0.0001, indicating that our estimates are nearly independent of the assumed initial progenitor mass. 

Some helium-burning reaction rates are notoriously uncertain and can significantly affect the oxygen mass fraction in WD cores \citep[e.g.,][]{degeronimo2017,pepper2022}, which in turn influences the $^{22}$Ne distillation threshold. However, the bottleneck for $^{22}$Ne production is the initial CNO content, which is largely insensitive to specific reaction rates. The resulting $^{22}$Ne mass fractions are closely related to the initial stellar metallicity, $Z$. This is true even in $\alpha$-enhanced metal-poor stars, since the total $^{14}$N, and thus the final $^{22}$Ne content, is governed by the total amount of CNO nuclei available in the star. Hence, our abundance estimates are robust against uncertainties in nuclear reaction rates.

\begin{figure}
\includegraphics[width=3.3in]{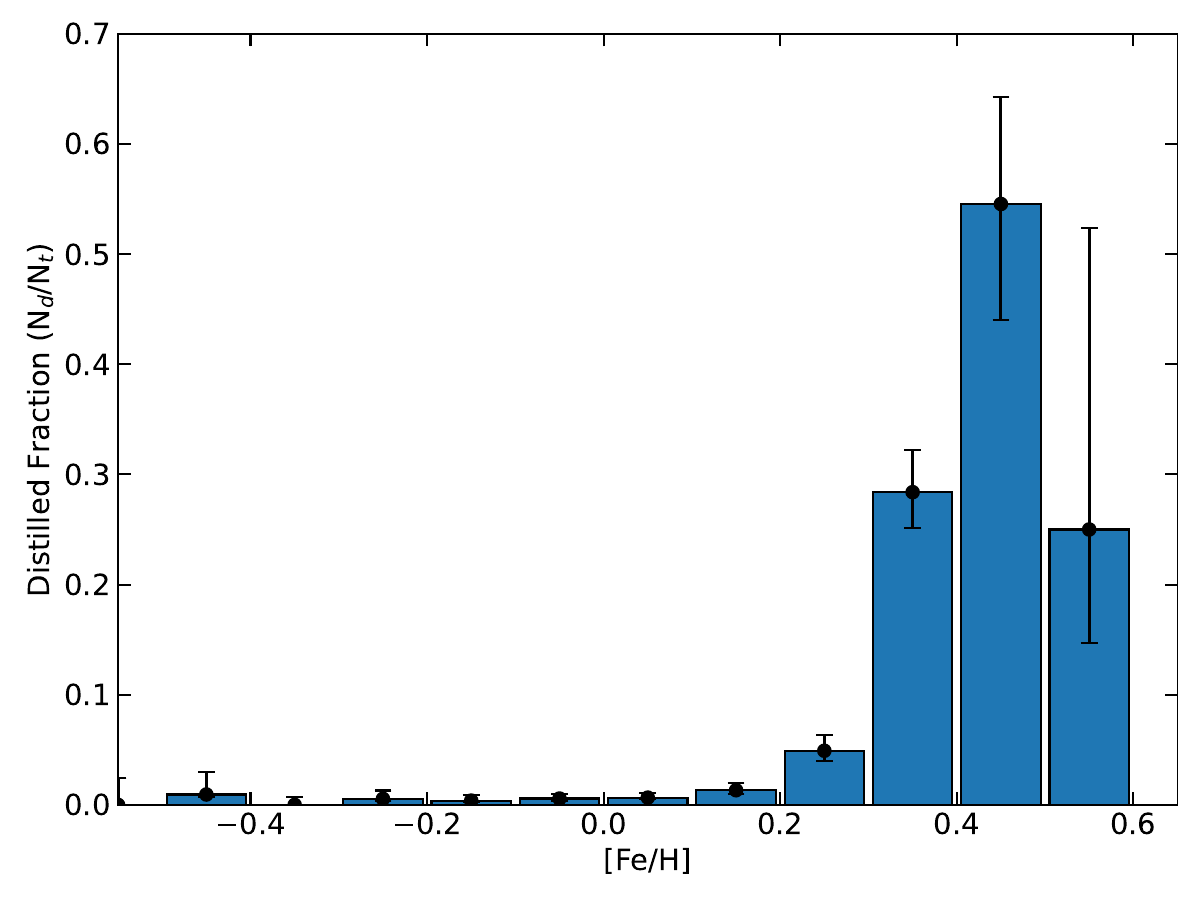}
\caption{Predicted fraction of distilled stars (N$_d$/N$_t$) as a function of metallicity for the white dwarf descendants of stars in the Hypatia Catalog \citep{hinkel14}. The metallicity axis is binned by 0.1 dex, and the error bars are calculated using a binomial distribution function. At the highest metallicities observed, up to $\sim50$\% of stars may leave behind a distilled white dwarf.}
\label{fig:2}
\end{figure}

\begin{figure*}
\includegraphics[width=7.1in]{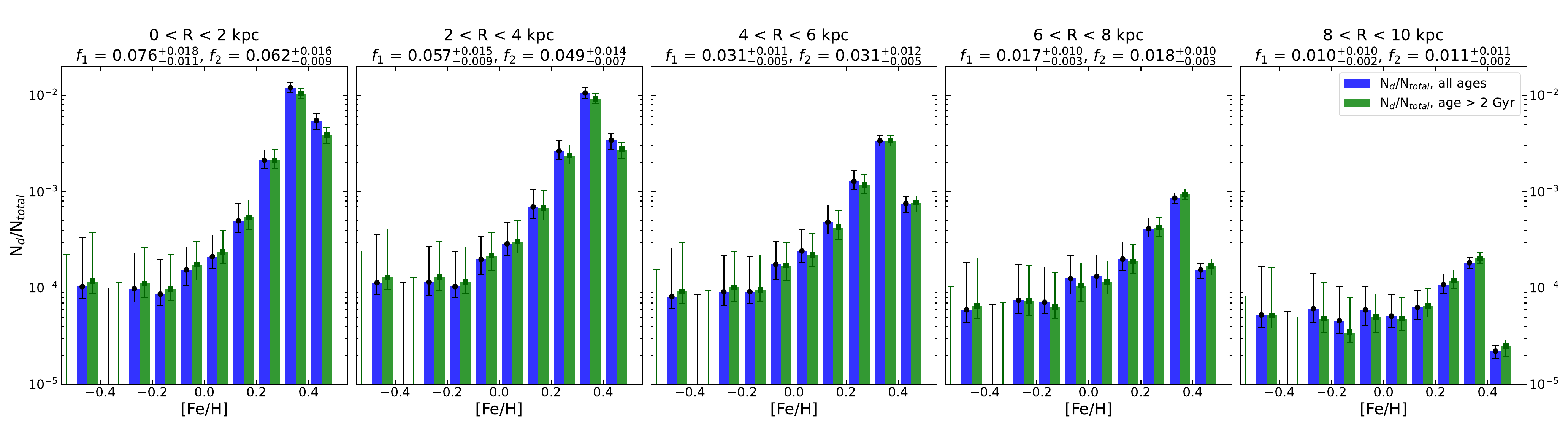}
\caption{Predicted number of distilled stars (normalized by the total number of stars in our models) as a function of metallicity for various Galactocentric distances. The blue and green bars show the results for stars of all ages and for those stars older than 2 Gyr, respectively. 
Each panel is labeled by the Galactocentric distance and the distilled fractions for stars of all ages (f$_1$) and older than 2 Gyr (f$_2$).}
\label{fig:3}
\end{figure*}

\section{Results}
\label{sec:3}

\subsection{Distilled White Dwarfs in the Solar Neighborhood}
\label{sec:3.1}

The top panel in Figure \ref{fig:1} shows the predicted $^{22}$Ne mass fraction for the WD descendants of the stars in the Hypatia Catalog. As expected, the overall trend is an increasing neon mass fraction with increasing metallicity, though there is some scatter in this relation. Figure~3 in \citet{blouin21} shows the conditions for distillation to occur. The key uncertainty is the oxygen mass fraction in the core. \citet{bauer2023} find central O mass fractions of $\approx$0.6-0.7 for WD models descended from MESA models of prior stellar evolution. Assuming X(O)=0.65, a neon mass fraction of X($^{22}$Ne) = 0.025 (marked by a dashed line in Figure~\ref{fig:1}) is the nominal threshold for triggering distillation at the center. However, only a small increase to X(O) = 0.7 (which is well within the range of current stellar evolution predictions; e.g. \citealt{A&C2022}) would be sufficient to increase the $^{22}$Ne threshold to $\sim3$\%.

More importantly, this threshold is the minimum abundance required for the distillation process to start at the WD's center, but it is unclear whether this minimum
abundance necessarily leads to multi-Gyr cooling delays. As discussed in \citet{bedard24} and \citet{salaris24}, $^{22}$Ne distillation
enriches the central layers at the expense of the outer layers, such that the latter eventually fall below the abundance threshold, which
may cause distillation to stop. The same authors show that multi-Gyr cooling delays should occur regardless of this uncertainty (whether
or not distillation continues beyond this point) provided that the $^{22}$Ne abundance is 3\% or higher, i.e. moderately higher than the
2.5\% threshold.

Out of the 4010 MESA calculations, 99 WDs exceed X($^{22}$Ne) = 0.025 implying that $2.5^{+1.1}_{-0.5}$\% of WDs in the solar neighborhood may form with sufficiently high $^{22}$Ne content to undergo distillation in the central layers. Alternatively,
for a higher $^{22}$Ne threshold for distillation, we find that $1.1^{+1.1}_{-0.2}$\% and $0.6^{+0.9}_{-0.2}$\% of WDs have
X($^{22}$Ne)$\geq0.0275$ and $\geq0.03$, respectively. Hence, depending on the exact threshold, 0.6 to 2.5\% of the WDs in the solar
neighborhood should go through distillation.

Within our sample, there are several outliers with large neon mass fractions at relatively low metallicities with [Fe/H] $\sim-2$. Figure~\ref{fig:1}, middle and bottom panels, show the C and N abundances of the Hypatia Catalog stars, respectively, along with the neon mass fraction shown as a color bar. BD+80 245 is the most C-rich star in the sample below [Fe/H] = $-1$, and both BD+80 245 and HD 19445 are chemically peculiar \citep[e.g., ][]{carney1997,reggiani23} and they are outliers in their N abundances. Hence, besides the few chemically peculiar stars enriched in CNO, halo and thick disk stars with [Fe/H] $<-1$ are not expected to go through distillation. 

Figure~\ref{fig:2} presents the fraction of WD progenitors predicted to undergo distillation (N$_d$/N$_t$), computed as a function of metallicity using the Hypatia Catalog and assuming the nominal threshold of X($^{22}$Ne) = 0.025. The data are binned in 0.1 dex intervals from [Fe/H] = $-1.5$ to $+1.0$, but only the relevant metallicity range is shown along with $\pm1 \sigma$ error bars. The distilled fraction exhibits a clear metallicity dependence, rising with increasing [Fe/H], and reaching a peak of $\sim$54\% around [Fe/H] $\sim +0.45$. This trend is consistent with the expected correlation between progenitor metallicity and $^{22}$Ne enrichment. 

For metallicity values above +0.5, the fraction drops to $\sim$25\%. However, this decrease is likely driven by small-number statistics rather than an intrinsic metallicity effect: only four stars in the Hypatia Catalog fall into this high-metallicity bin, making the estimate highly uncertain, as shown by the error bars. At lower metallicities ([Fe/H]$< -0.6$), the distilled fraction is effectively zero, confirming that low-metallicity stars typically do not produce WDs that undergo distillation, unless they are chemically peculiar stars that are enriched in CNO.

While the progenitor stars in our sample produce standard-mass C/O core WDs ($ 0.5 \leq M/M_{\odot} < 1.1$), and thus cannot explain the origin of the massive distilled WDs on the Q-branch \citep[see][]{cheng19,bedard24}, this subset represents the majority of the WD population in the Galaxy. For these average-mass WDs, our results yield a distilled fraction between 0.6\% and 2.5\%, depending on the adopted $^{22}$Ne threshold. Therefore, WD remnants of metal-rich progenitors can explain at least part of the observed pile-up in the 100 pc WD sample \citep{kilic2025}, though the exact impact depends on the true abundance required to sustain multi-Gyr cooling delays.

The spectroscopic abundances in the Hypatia Catalog reflect present-day photospheric compositions, which are likely lower than the initial bulk metallicity of stars due to gravitational settling and element diffusion over time. These processes can reduce photospheric abundances of heavy elements by approximately 10–20\% relative to their primordial values \citep{Lodders2020, hinkel2022}. Although the relative ratios among heavy elements remain constant, the total CNO content estimate from surface measurements likely represents a lower bound on the progenitor's true $^{14}$N and thus $^{22}$Ne content. Hence, the distilled WD fraction may be even higher than our estimate.

\begin{figure}
\includegraphics[width=3.3in]{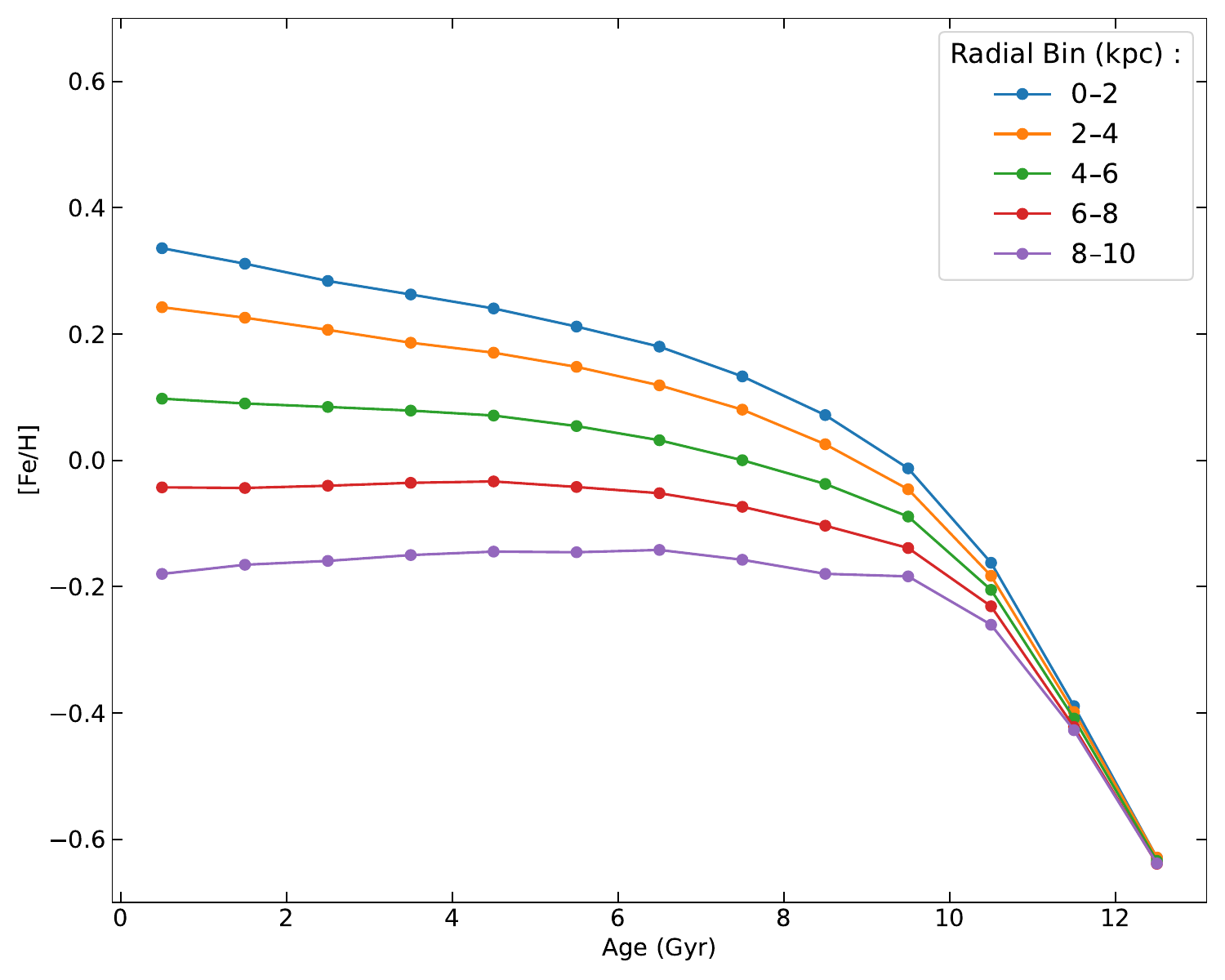}
\caption{Age-Metallicity relationship for the stars in the Galactic Stellar Population Model of \cite{s&h2021} for the different radial bins. The points indicate the average [Fe/H] for every 1 Gyr bin. The average metallicity value is relatively flat for ages younger than 8 Gyr in the outer disk (green, red, purple lines), but it is slightly steeper in the inner Milky Way (blue and orange lines).}
\label{fig:4}
\end{figure}

\subsection{Distillation throughout the Galaxy}
\label{sec:3.2}

To explore how the fraction of distilled stars changes across the Milky Way, we adopted the Galactic stellar population model developed by \cite{s&h2021}, which combines parametric star formation histories with observationally-constrained chemical evolution. The model's chemistry is calibrated by the APOGEE DR14 \citep{apogeeDR14} and by GALAH DR3 \citep{galahDR3} surveys, and the velocity dispersion is constrained by GALAH and LAMOST survey information \citep{sharma2021}. It also includes vertical and radial stellar density profiles, making it well-suited to capture the chemical structure of the Milky Way disk. 

Stars in the model are generated according to a universal initial mass function (independent of age and metallicity of the interstellar medium). Thus, the metallicity distribution of present-day WD progenitors mirrors that of all stars ever formed. With this framework, we aim to assess where in the Galaxy WDs are most likely to experience distillation and, consequently, where long-lived WD habitable zones might arise. We assume that each star is equally likely to end up as a WD, ignoring the dependence of the main-sequence lifetime on the initial mass.

Since the Galactic model provides metallicity information but not explicit $^{22}$Ne abundances, we mapped the empirical relation between [Fe/H] and distillation fraction derived from the Hypatia Catalog (see Section~\ref{sec:3.1}) onto the model’s [Fe/H] distributions. Assuming this relation holds throughout the Galaxy, we then applied the Hypatia-based distilled fractions to each metallicity bin across different Galactocentric radii, which allows us to estimate the distribution of distilled WDs in the Milky Way. In particular, we decided to separate the sample into 2 kpc bins, spanning Galactocentric distances from 0 to 10 kpc. To ensure consistency with the local sample, we further restricted the analysis to stars located within $|z|<$0.3 kpc of the Galactic plane.

\begin{table*}
    \centering
    \def\arraystretch{1.50}
    \setlength{\tabcolsep}{5.0pt}
    \caption{Distilled Stars Percentage for Different $^{22}$Ne Concentrations in the Hypatia Catalog and the Galactic Model for different Galactocentric radii. Uncertainties are $1\sigma$ binomial errors.}
    \begin{tabular}{l||c||c|c|c|c|c}
   $X(^{22}$Ne) &  Hypatia (\%) & \multicolumn{5}{c}{Galactic Model (\%)} \\
   \cline{3-7}
    &  & 0--2 kpc & 2--4 kpc & 4--6 kpc & 6--8 kpc & 8--10 kpc \\
       \hline\hline
        0.0250 & 2.5$^{+1.1}_{-0.5}$ & 7.6$^{+1.8}_{-1.1}$ & 5.7$^{+1.5}_{-0.9}$ & 3.1$^{+1.1}_{-0.5}$ & 1.7$^{+1.0}_{-0.3}$ & 1.0$^{+1.0}_{-0.2}$\\
        0.0275 & 1.1$^{+1.1}_{-0.2}$ & 3.7$^{+1.6}_{-0.8}$ & 2.6$^{+1.3}_{-0.3}$ & 1.4$^{+1.0}_{-0.3}$ & 0.8$^{+0.8}_{-0.2}$ & 0.5$^{+0.9}_{-0.1}$ \\
        0.0300 & 0.6$^{+0.9}_{-0.2}$ & 1.8$^{+1.4}_{-0.5}$ & 1.3$^{+1.2}_{-0.4}$ & 0.7$^{+0.9}_{-0.2}$ & 0.4$^{+0.8}_{-0.1}$ & 0.2$^{+0.8}_{-0.1}$ \\
       \hline\hline
    \end{tabular}
    \label{tab:1}
      \vspace{3mm}
\end{table*}

Figure~\ref{fig:3} presents the resulting histograms of WD progenitors expected to undergo distillation (N$_d$) normalized by the total number of stars in our model (N$_{\rm total}$) as a function of metallicity across five Galactocentric radial bins from 0 to 10 kpc. The blue bars indicate the normalized number of WD progenitors of all ages that will go through distillation using the X($^{22}$Ne) = 0.025 threshold, while the green bars show the results when restricting the sample to WD progenitor stars older than 2 Gyr, which helps us reproduce an older metallicity profile. This age cut is motivated by the fact that distillation can only begin after WDs' core crystallization, which typically occurs after $\sim$2 Gyr of cooling for a 0.6 M$_\odot$ WD. Thus, the green bars reflect more accurately the overall distilled WD population today (and not just the massive WDs that crystallize relatively quickly).

As expected, the population of distilled WDs exhibits a strong radial gradient: the highest fractions are found in the inner regions, with a peak of 7.6\% inside 2 kpc and 5.7\% between 2–4 kpc. Beyond this, the distilled fraction declines steadily to 3.1\%, 1.7\%, and 1.0\% in the 4–6 kpc, 6–8 kpc, and 8–10 kpc bins, respectively. Focusing on the solar neighborhood with 7.9 $<$ R $<$ 8.5 kpc and $|z| < 0.3$ kpc, we find a distilled fraction of 1.1$^{+0.9}_{-0.2}$\%. This is lower than the 2.5\% distilled fraction estimated from the Hypatia Catalog sample above, but the $1\sigma$ upper limit of 2\% is comparable to that estimate. Similarly, when considering only stars older than 2 Gyr, the distilled fractions are slightly lower, with the most pronounced reductions occurring in the inner Galaxy. In this older subset, we observe distilled fractions of 6.2\% and 4.9\% (0–2 kpc and 2–4 kpc, respectively), while fractions beyond 4 kpc remain relatively unchanged. 

This behavior reflects the underlying metallicity gradient with age in the Galaxy. Metal-rich stars, which dominate the inner Galaxy, are usually younger and more likely to exceed the X($^{22}$Ne) threshold required for distillation. In contrast, older stars tend to be more metal-poor, producing less $^{22}$Ne, and therefore no distillation happens. Figure~\ref{fig:4} further illustrates this chemical-age dependence by showing the [Fe/H]–age relationships of stars in the \cite{s&h2021} Galaxy model for the different radial bins. The average metallicity value is relatively flat for the past 8 Gyr in the outer disk \citep[see also Figure 1 in][]{s&h2021}. Whereas, the trend is slightly steeper in the inner Galaxy, likely because of the role of radial migration in the Galactic model: older metal-poor stars originating in the outer disk tend to migrate inwards, lowering the average [Fe/H] of older populations near the Galactic center.

Table~\ref{tab:1} presents the fraction of distilled WDs as a function of Galactocentric distance for various values of
the X($^{22}$Ne) threshold. The table also includes the results for the Hypatia Catalog sample for reference. Using the Galactic
model, we find that the distilled WD fraction in the innermost bin (R$<2$ kpc) drops from 7.6\% in the optimistic case to just 1.8\% when
adopting the $X(^{22}{\rm Ne})=0.03$ threshold. Overall, using the higher threshold of 3\% neon mass fraction lowers the number of
distilled WDs by about a factor of four. These values provide a conservative range that reflects the uncertainties in both the core
composition and the detailed physics of the distillation process.

Regardless of these uncertainties, our results indicate that WD progenitors with sufficiently high $^{22}$Ne content are preferentially located in the inner disk of the Galaxy. As a result, this region is expected to host the majority of WDs that undergo distillation, which implies that long-lived WDs' habitable zones are most likely to occur in the inner Milky Way.

\section{Conclusions}
\label{sec:4} 

To determine the population of WDs predicted to go through $^{22}$Ne distillation in the solar neighborhood, we take advantage of the CNO abundances and [Fe/H] for $\sim$4000 progenitor stars from the Hypatia Catalog. Assuming a 3 M$_{\odot}$ progenitor mass, we used MESA simulations to estimate the $^{22}$Ne mass fraction, X($^{22}$Ne), in their WD remnants. In our baseline case with a threshold X($^{22}$Ne)= 0.025, we find that 2.5\% of the population of WDs in the Solar Neighborhood exceeds this value and would thus trigger the distillation process and experience a multi-Gyr cooling delay. The strong observed correlation between metallicity and $^{22}$Ne content suggests that single-star evolution of metal-rich progenitors can naturally account for a non-negligible fraction of standard-mass C/O core WDs with considerable cooling delays and long-lived habitable zones.

We extended this analysis to the Galactic scale using the stellar population model of \cite{s&h2021}. For the baseline threshold, the predicted distilled fraction decreases monotonically with Galactocentric radius, from 7.6$^{+1.8}_{-1.1}$\% near the Galactic center ($R < 2$ kpc) to 1.0$^{+1.0}_{-0.2}$\% in the outer disk ($8 < R < 10$ kpc). This radial trend holds even when restricting the sample to stars older than 2 Gyr, which reproduces an older metallicity profile that more accurately reflects the WD population today. 

When accounting for higher $^{22}$Ne thresholds of 2.75\% and 3.0\%, these fractions are reduced by roughly a factor of four. For instance, the distilled fraction drops from 2.5\% to 0.6\% for the solar neighborhood and from 7.6\% to 1.8\% for the inner Galaxy. These highlight the sensitivity of the predicted distilled WD population to the assumed abundance threshold. Nonetheless, our results indicate that single-star evolution can explain up to a few percent of WDs undergoing distillation locally, with higher fractions in the inner Galaxy where long-lived habitable zones around WDs are more common.

\section{Acknowledgments}
This work is supported in part by the NSF under grant AST-2205736 and the NASA under grants 80NSSC22K0479, 80NSSC24K0380, and 80NSSC24K0436.

The research shown here acknowledges use of the Hypatia Catalog Database, an online compilation of stellar abundance data as described in \citet{hinkel14}, which was supported by NASA's Nexus for Exoplanet System Science (NExSS) research coordination network and the Vanderbilt Initiative in Data-Intensive Astrophysics (VIDA).


\bibliography{Distillation_paper}{}
\bibliographystyle{aasjournal}

\end{document}